\documentclass[aps,twocolumn,amsmath,amssymb,showpacs]{revtex4-1}
\usepackage{graphicx,color,hyperref}
\begin{document}
\def\be{\begin{equation}}
\def\ee{\end{equation}}
\def\bea{\begin{eqnarray}}
\def\eea{\end{eqnarray}}
\def\bef{\begin{figure}[h!]}
\def\eef{\end{figure}}

\def\a{\alpha}
\def\th{\theta}
\def\o{\omega}
\def\eps{\epsilon}
\def\fr{\frac}
\def\l{\label}

\title{One-dimensional lattice of oscillators coupled through power-law interactions: Continuum
limit and dynamics of
spatial Fourier modes}
\author{Shamik Gupta,$^1$ Max Potters,$^{1,2}$ Stefano Ruffo$^{1,3}$}
\affiliation{$^1$Laboratoire de Physique de l'\'{E}cole Normale Sup\'{e}rieure de Lyon, Universit\'{e}
de Lyon, CNRS, 46 All\'{e}e d'Italie, 69364 Lyon c\'{e}dex 07, France\\
$^2$Delft Center for Systems and Control, Technical University Delft,
Mekelweg 2, 2628 CD Delft, The Netherlands \\
$^3$Dipartimento di Energetica ``Sergio Stecco'' and CSDC,
Universit\`{a} di Firenze, CNISM and INFN, via S. Marta 3, 50139 Firenze, Italy
}
\begin{abstract}
We study synchronization in a system of phase-only oscillators
residing on the sites of a one-dimensional periodic lattice. The
oscillators interact with a
strength that decays as a power law of the separation along the lattice
length and is normalized by a size-dependent constant.  The exponent $\a$ of the
power law is taken in the range $0 \le \a <1$. The oscillator frequency
distribution is symmetric about its mean (taken to be zero), and is non-increasing on $[0,\infty)$.
 In the
continuum limit, the local density of oscillators evolves in time
following the continuity equation that expresses the conservation of the
number of oscillators of each frequency under the dynamics. This equation admits as a stationary
solution the unsynchronized state uniform both in phase and over the space of
the lattice. We perform a linear stability analysis of this state to show that when it is
unstable, different spatial Fourier modes of fluctuations have different
stability thresholds beyond which they grow exponentially in time with
rates that depend on the Fourier modes. However, numerical simulations
show that at long times, all the
non-zero Fourier modes decay in time, while only the zero Fourier mode
(i.e., the ``mean-field" mode) grows in time, thereby dominating the
instability process and driving
the system to a synchronized state. Our
theoretical analysis is supported by extensive numerical simulations.
\end{abstract}
\date{\today}
\maketitle
\section{Introduction: The model}
A large collection of coupled oscillating components
of different natural frequencies getting spontaneously synchronized to oscillate
at a common frequency is a phenomenon commonly observed in a variety of
biological and physical systems, e.g., yeast cell suspensions \cite{Ghosh:1971,Aldridge:1976}, networks of pacemaker cells in the
heart \cite{Peskin:1975,Michaels:1987}, groups of flashing fireflies
\cite{Buck:1976,Buck:1988}, arrays of superconducting
Josephson junctions \cite{Wiesenfeld:1996,Wiesenfeld:1998}, and many
others \cite{Pikovsky:2003}.

The Kuramoto model provides a simple setting to investigate the physics
of synchronization \cite{Kuramoto:1984,Strogatz:2000,Acebron:2005}. The model considers a large population of phase-only oscillators
of distributed natural frequencies. The oscillators are globally coupled with
coupling strengths that are equal and independent of their spatial
distribution. The governing dynamical
equations are
\be
\fr{d\th_i}{dt}=\o_i+\fr{K}{N}\sum_{j=1,j \ne i}^N\sin(\th_j-\th_i),
\l{timeevolution}
\ee
for $i=1,2,\ldots,N \gg 1$. Here, $\th_i$ is a periodic variable of period
$2\pi$ that denotes the phase of the $i$th
oscillator whose natural frequency is given by $\o_i \in R$. The parameter $K \ge 0$
denotes the coupling strength. The frequencies are distributed according
to a probability density $g(\o)$, assumed to be one-humped and symmetric
about its mean. It is known that above a critical coupling strength
$K_c=2/(\pi g(0))$, the stationary state that the system attains at long
times is a synchronized state. For $K < K_c$, however, there is no synchronization at long times,
and each oscillator keeps running at its own natural
frequency. 

Although the Kuramoto model has been studied extensively in recent years, much less is 
known about the case in which the oscillators are
coupled with a strength that is a function of their spatial
distribution, e.g., an inverse power law in the separation. Such a form
of interaction is relevant for many situations, for example, in the study of long-range synchrony in
a network of coupled neurons \cite{Konig:1995} or in modeling flashing fireflies where the intensity of light signals
carrying information from one firefly to another falls off in the three-dimensional space as the
inverse of square of the distance from the source. 

A simple modification of the Kuramoto model (\ref{timeevolution}) to
include a power-law interaction is to consider a one-dimensional periodic lattice of
$N$ sites, with each site containing one oscillator, and with the
oscillators interacting with one another with a strength that decays as a power law of the
separation along the lattice length \cite{Rogers:1996}. Consequently, the governing
equations are modified from (\ref{timeevolution}) to 
\be
\fr{d\th_{i}}{dt}=\omega_{i}+\frac{K}{\widetilde{N}}\sum_{j=1,j \ne
i}^{N}\fr{\sin(\th_j-\th_i)}{|x_j-x_i|^{\a}}; ~~\a \ge 0.
\l{EOM}
\ee
Here, $x_j=j\eps$ is the coordinate of the $j$th site on the lattice,
where $\eps$ is the lattice constant. The constant $\widetilde{N}=\sum_{j=1}^N|x_{j}-x_{i}|^{-\alpha} ~\forall ~i$ is a
size-dependent normalization. Being
on a periodic lattice, we adopt the closest distance convention:
\be
|x_j-x_i|={\rm min}(|x_j-x_i|,1-|x_j-x_i|),
\ee
where the total length $N\eps$ of the lattice has been taken to equal
unity without loss of generality. From Eq. (\ref{EOM}), we see that for $\a < 1$, the cumulative interaction of one oscillator with all the
remaining oscillators with aligned phases would diverge in the continuum limit $N
\to \infty$ in the absence of the normalization
$\widetilde{N}$, which explains its inclusion
\cite{Rogers:1996,Massunaga:2002}. Note that the case $\a=0$ of the
model corresponds to the usual Kuramoto model (\ref{timeevolution}). Let
us mention that a model with space-dependent phases as in Eq. (\ref{EOM}), but
with a different form of coupling between the oscillators has recently been studied in Ref.
\cite{Pikovsky:2011}. 

A previous study of the model (\ref{EOM}) addressed by numerical simulations the issue of
synchronization, in particular, the conditions for its emergence
at long times as a function of $\a$ for a fixed value of $K$ \cite{Rogers:1996}. It was reported that a critical
value $\a_c=2$ exists, such that only when $\a < \a_c$ does the system in the limit $N \to \infty$
synchronize at a finite value of $K$. For $\a > \a_c$, there is no
finite value of $K$ at which synchronization occurs. A recent 
theoretical study however predicted the value $3/2$ for $\alpha_{c}$ \cite{Chowdhury:2010}. In Ref. 
\cite{Marodi:2002}, the model was studied without the normalization
$\widetilde{N}$, and for finite values of
$N$. In this work, mainly numerical simulations were employed to
investigate the dependence of $N$ on $\a$ and $K$ that leads to 
synchronization.

In this paper, we consider the model (\ref{EOM}) with values of $\a$ in the range $0
\le \a <1$. We take the frequency distribution $g(\o)$ to be symmetric about its mean (taken to
be zero) and non-increasing on $[0,\infty)$. We analyze the model
(\ref{EOM}) in the continuum limit $N \to \infty$, when the lattice is replaced by a continuous line characterized by the
spatial coordinate $x \in [0,1]$. Since in this limit each infinitesimal element
$dx$ contains a diverging number of oscillators, it is natural to define a local density of
oscillators $\rho(\th;\o, x,t)dx$ such that of all oscillators with
natural frequency $\o$ that are located between $x$ and $x+dx$ at time
$t$, the fraction $\rho(\th;\o,x,t)d\th dx$ have their phase between $\th$ and $\th+d\th$. 
Because the dynamics conserves the total number of oscillators with
frequency $\o$, the time evolution of $\rho(\th; \o,x,t)$ is given by
the continuity equation. In this equation, the local
densities at different spatial locations appear coupled due to the
interaction between the oscillators.

The unsynchronized state, uniform both in the phase $\th$ and in the spatial coordinate
$x$, solves the continuity equation in the stationary state. By
performing a linear stability analysis of such a state, we show that when it is
unstable, different spatial Fourier modes of fluctuations have different
stability thresholds beyond which they grow exponentially in
time with different rates. However, numerical simulations
show that at long times, all the non-zero
Fourier modes decay in time, while the zero mode (the ``mean-field"
mode) grows in time and dominates the instability
process, thereby driving
the system to a synchronized state.
Such a long-time dominance of the 
mean-field mode in characterizing dynamic instability has been observed in systems with power-law interactions
evolving under a deterministic Hamiltonian dynamics
\cite{Bachelard:2011}. The present study demonstrates this
dominance for the dissipative dynamics of the model (\ref{EOM}). Our
theoretical predictions for the growth rates of the unstable modes are corroborated by
numerical simulations.  

The outline of this paper is as follows. In the following Section, we
consider the model (\ref{EOM}) in the continuum limit $N \to \infty$, and analyze
the linear stability of the unsynchronized state. In the region in which it
is unstable, we derive analytic expressions for the stability thresholds
and the growth rates of the various Fourier modes of fluctuations. In Section
\ref{simulations}, we test our theoretical predictions for the growth
rates by performing numerical simulations, and illustrate the
dominance of the mean-field mode in characterizing the long-time
instability of the unsynchronized state. In
simulations, we employ a fast numerical algorithm to compute the
interaction among the oscillators, the details of which are relegated to
the Appendix \ref{app1}. In the final Section, we draw our conclusions.
 
\section{Continuum limit analysis}
In this section, we consider the model (\ref{EOM}) in the continuum limit $N \to
\infty$, and investigate in detail the issue of linear stability of the
unsynchronized state. Finite-$N$ effects are known to be quite subtle and difficult to tackle even
for the usual Kuramoto model \cite{Strogatz:2000,Acebron:2005}, so we do
not attempt a finite-$N$ analysis of the model (\ref{EOM}) in
this work.

In the continuum limit, the lattice of the system (\ref{EOM}) is densely filled with sites.
Let the continuous variable $x \in [0,1]$ stand for the spatial
coordinate along the lattice length. Now that each infinitesimal element $dx$ contains
a diverging number of oscillators, we define a local density of
oscillators $\rho(\th;\o, x,t)dx$ such that of all oscillators with
natural frequency $\o$ that are located between $x$ and $x+dx$ at time
$t$, the fraction $\rho(\th;\o,x,t)d\th dx$ have their phase between $\th$ and $\th+d\th$. This
density is non-negative, $2\pi$-periodic in $\th$, and obeys the
normalization
\be
\int_0^{2\pi}d\th ~\rho(\theta;\omega,x,t)=1.
\l{rho-normalization}
\ee
The equations of motion (\ref{EOM}) become
\bea
&&\hspace{-0.5cm}\fr{\partial \theta(\o,x,t)}{\partial t}=\nonumber \\
&&\hspace{-0.5cm}\o+\kappa(\a)K\int d\th'd\o'dx'\fr{\sin(\th'-\th)}{|x'-x|^{\alpha}}\rho(\th';\o',x',t)
g(\o'), 
\l{EOMcontinuum}
\eea
where $\kappa(\a)$ is such that $N/\widetilde{N} \to \kappa(\a)$ as $N
\to \infty$, and therefore, one has 
\be
\kappa(\a)^{-1}=\int_{x-1/2}^{x+1/2}\fr{dx'}{|x'-x|^\a}.
\l{kappa}
\ee
Since we consider $\a$ in the range $0 \le \a < 1$, the above integral is convergent.
The number of oscillators with frequency $\o$ is conserved by the
dynamics, so that the time evolution of $\rho(\th; \o,x,t)$ follows the continuity equation
\be
\fr{\partial\rho}{\partial
t}=-\fr{\partial}{\partial\th}\Big(\rho\fr{\partial \theta}{\partial
t}\Big),
\l{continuity-equation}
\ee
where $\fr{\partial \theta}{\partial t}$ is given by Eq.
(\ref{EOMcontinuum}).

Next, consider the unsynchronized state uniform both in the phase $\th$ and in the
spatial coordinate $x$:
\be
\rho_{0}(\th;\o,x,t)=\fr{1}{2\pi}.
\l{incoherent-state}
\ee
Such a state is evidently a stationary solution of the time evolution
(\ref{continuity-equation}). To study its linear stability under the
dynamics, consider adding small fluctuations, so that
\be
\rho(\th;\o,x,t)=\fr{1}{2\pi}+\delta\rho(\th;\o,x,t);
~~\delta\rho(\th;\o,x,t) \ll 1.
\l{delta-rho-definition}
\ee
Substituting into Eq. (\ref{continuity-equation}) and keeping terms to
linear order in $\delta \rho$, we find that $\delta \rho$ satisfies
\bea
&&\fr{\partial\delta\rho(\th;\o,x,t)}{\partial
t}=-\omega\frac{\partial\delta\rho(\th;\o,x,t)}{\partial\theta}\nonumber \\
&&+\fr{\kappa(\alpha)K}{2\pi}\int d\th'd\o'dx'
~\fr{\cos(\th'-\th)}{|x'-x|^{\alpha}}\delta\rho(\th';\o',x',t)g(\o').
\nonumber \\
\l{linearized-equation}
\eea
Expressing $\delta \rho$ in terms of its Fourier series with respect
to the periodic variable $\th$ as
\be
\delta\rho(\th;\o,x,t)=
\sum_{k=-\infty}^\infty\widehat{\delta\rho}_k(\o,x,t) e^{ik\th},
\l{delta-rho-definition1}
\ee
we find from Eq. (\ref{linearized-equation}) that only the modes $k=\pm
1$ are affected by the coupling between the oscillators, and that $\widehat{\delta \rho}_{\pm
1}$ satisfies
\bea
&&\fr{\partial\widehat{\delta\rho}_{\pm1}(\o,x,t)}{\partial
t}=\mp i\o\widehat{\delta\rho}_{\pm1}(\o,x,t)\nonumber \\
&&+\fr{\kappa(\a)K}{2}\int d\omega' dx'
\fr{\widehat{\delta\rho}_{\pm1}(\omega',x',t)}{|x'-x|^{\a}}g(\o').
\l{linearized-equation-1}
\eea
One thus gets a set of equations for each position $x$, all coupled
together by the second term on the right.

Since we have a periodic lattice, it is natural to consider the spatial
Fourier series of $\widehat{\delta \rho}_{\pm 1}(\o,x,t)$:
\be
\widehat{\delta \rho}_{\pm 1}(\o,x,t)=\sum_{m=-\infty}^\infty \overline{\delta \rho}_{\pm
1,m}(\o,t)e^{i2\pi mx}.
\l{spatial-FT}
\ee
On substituting in Eq. (\ref{linearized-equation-1}), we get 
\bea
&&\fr{\partial\overline{\delta\rho}_{\pm1,m}(\o,t)}{\partial
t}=\mp i\o\overline{\delta\rho}_{\pm1,m}(\o,t)\nonumber \\
&&+\fr{\kappa(\a)K\Lambda_m(\a)}{2}\int d\omega'~ 
\overline{\delta\rho}_{\pm1,m}(\omega',t)g(\o'),
\l{linearized-equation-2}
\eea
where
\bea
\Lambda_m(\a)&=&\int_{x-1/2}^{x+1/2} dx' ~\fr{e^{i2\pi
m(x'-x)}}{|x'-x|^\a}.
\l{lambda}
\eea

Note that $\Lambda_m(\a)=\Lambda_{-m}(\a)$. We have checked numerically that
$\Lambda_m(\a) \ge 0$, and that it is a monotonically decreasing
function of $m$, see Fig. \ref{fig0}. These properties can also be
proved analytically \cite{Campa-note}. It is also easy to prove that
$\Lambda_m(\a) \to 0$ as $m \to \infty$. Defining $w\equiv my$, we get
\be
\Lambda_m(\a)=\fr{2}{m^{1-\a}}\int_0^{m/2} dw ~\fr{\cos(2\pi w)}{w^\a}.
\l{lambda2}
\ee
In the limit $m \to \infty$, the integral evaluates to a finite constant equal
to $(2\pi)^{\a-1} \sin (\pi \a/2) \Gamma(1-\a)$, while the prefactor
tends to zero, yielding
\be
\lim_{m \to \infty} \Lambda_m(\a)=0.
\ee

\begin{figure}[h!]
\includegraphics[width=80mm]{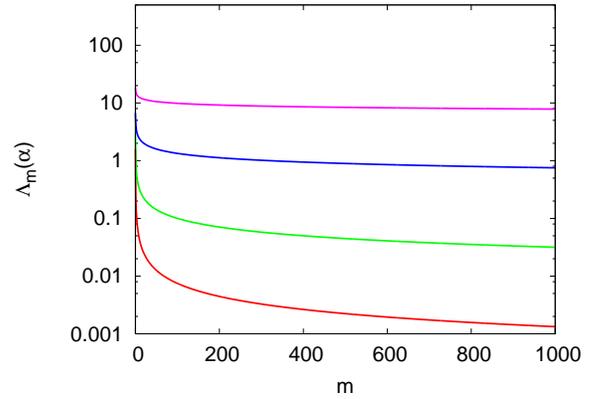}
\caption{(Color online) $\Lambda_m(\a)$ as a function of $m$, obtained
by numerically evaluating the integral in Eq. (\ref{lambda}). The
different curves correspond to different $\a$ values given by
$\a=0.25,0.5,0.75,0.9$ (bottom to top).}
\l{fig0}
\end{figure}

Following \cite{Strogatz:2000}, we consider the
right hand side of Eq. (\ref{linearized-equation-2}) to
define a linear operator $A$ as
\bea
&&A\overline{\delta\rho}_{\pm1,m}(\o,t)=\nonumber \\
&&\mp
i\o\overline{\delta\rho}_{\pm1,m}(\o,t)+\fr{\kappa(\a)K\Lambda_m(\a)}{2}\int
d\omega'~
\overline{\delta\rho}_{\pm1,m}(\omega',t)g(\o'). \nonumber \\
\l{operatorA}
\eea
Then, in terms of the eigenvalues $\lambda_m$ of the operator $A$, Eq.
(\ref{linearized-equation-2}) gives
\be
\overline{\delta \rho}_{\pm 1,m}(\o,t)= \widetilde{\delta \rho}_{\pm
1,m}(\o,\lambda_m)e^{\lambda_m t},
\l{temporal-LT}
\ee
with $\widetilde{\delta \rho}_{\pm
1,m}$ satisfying
\bea
&&\widetilde{\delta\rho}_{\pm1,m}(\o,\lambda_m)=\fr{\kappa(\a)K\Lambda_m(\a)}{2(\lambda_m
\pm i\o)}\int d\omega' 
\widetilde{\delta\rho}_{\pm1,m}(\omega',\lambda_m)g(\o'). \nonumber \\
\l{linearized-equation-3}
\eea
Multiplying both sides of Eq. (\ref{linearized-equation-3}) by $g(\o)$,
and then integrating with respect to $\o$, one gets
\be
I_{\pm, m}(1-J_{\pm,m})=0,
\l{dispersion-1}
\ee
where
\bea
I_{\pm, m}&=&\int_{-\infty}^\infty d\o ~\widetilde{\delta \rho}_{\pm 1,m}(\o,\lambda_m)g(\o), \\
J_{\pm, m}&=&\fr{\kappa(\a) K\Lambda_m(\a)}{2}\int_{-\infty}^\infty d\o
~\fr{g(\o)}{(\lambda_m \pm i\o)}.
\eea
Since $I_{\pm,m} \ne 0$, Eq. (\ref{dispersion-1}) gives the dispersion relation
\be
J_{\pm, m} =1.
\l{dispersion}
\ee

Let us consider $g(\o)$ to be symmetric about its mean. We may assume a
zero mean without loss of generality (one can easily achieve this by going into a rotating frame of
reference). Moreover, we take $g(\o)$ to be non-increasing
on $[0,\infty)$, i.e., $g(\o) \ge g(\nu)$ whenever $\o \le \nu$.
Examples include common
frequency distributions of interest like the Gaussian and the Lorentzian distributions. One can prove by using Theorem $2$ in Ref.
\cite{Strogatz:1990} that for such a $g(\o)$, Eq.
(\ref{dispersion}) has at most one solution for $\lambda_m$, and if it
exists, it is necessarily real. Then, Eq. (\ref{dispersion}) gives
\be
\kappa(\a) K \Lambda_m(\a) \int_0^\infty
d\o ~\fr{\lambda_m}{\lambda_m^2+\o^2}g(\o)=1.
\l{final-dispersion}
\ee
The above equation implies that $\lambda_m \ge 0$, as otherwise the
left hand side is negative. Since the $m$th spatial mode of fluctuations $\overline{\delta
\rho}_{\pm 1,m}$ has the time dependence $\overline{\delta \rho}_{\pm 1,m} \sim e^{\lambda_m t}$ (see Eq.
(\ref{temporal-LT})), it follows
that this mode is either linearly neutrally stable corresponding to
$\lambda_m =0$, or that it is linearly unstable corresponding to
$\lambda_m >0$. The borderline between these two behaviors is achieved
at the critical coupling $K_c^{(m)}$, obtained by taking the limit $\lambda_m
\to 0^+$ in Eq. (\ref{final-dispersion}); we get 
\be
K_c^{(m)}=\fr{2}{\kappa(\a) \Lambda_m(\a)\pi g(0)}.
\l{Kcm}
\ee
Combining the last equation with Eq. (\ref{final-dispersion}), the growth
rate $\lambda_m >0 $ for $K > K_c^{(m)}$ is given by
\be
\fr{2K}{\pi g(0)K_c^{(m)}} \int_0^\infty
d\o ~\fr{\lambda_m}{\lambda_m^2+\o^2}g(\o)=1.
\l{final-dispersion1}
\ee

With $\Lambda_m(\a)=\Lambda_{-m}(\a)$, it follows that the Fourier modes
$m$ and $-m$ have the same critical thresholds and the same growth rates. 
Since $\Lambda_{|m|}(\a)$ decreases on increasing $|m|$, we conclude that
\be
K_c^{(0)} < K_c^{(1)} < K_c^{(2)} < \ldots.
\l{K-hierarchy}
\ee
The unsynchronized state (\ref{incoherent-state}) is linearly stable if all
the spatial modes of fluctuations decay in time; using Eq.
(\ref{K-hierarchy}), this is achieved when $K < K_c^{(0)}$.
On the other hand, it becomes unstable for $K > K_c^{(0)}$. 

For $\a=0$, when our model (\ref{EOM}) reduces to the Kuramoto model
(\ref{timeevolution}), Eqs. (\ref{kappa}) and (\ref{lambda}) give
$\kappa(\a)\Lambda_m(\a)=\delta_{m,0}$. In this limit, the
oscillators are globally coupled with equal coupling strengths, and
therefore, it is physical to talk of only the zero Fourier mode. This
mode has the critical
coupling given by Eq. (\ref{Kcm}) to be $K_c^{(0)}=2/\pi g(0)$, such that it
is linearly unstable for higher values of $K$; this matches with known
results for the Kuramoto model \cite{Strogatz:2000,Acebron:2005}. For
finite values of $K$, all
the other modes are linearly neutrally stable because of diverging
critical thresholds. 

\begin{figure}[h!]
\includegraphics[width=80mm]{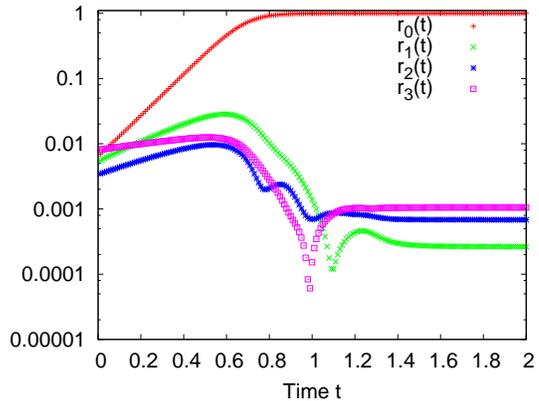}
\caption{(Color online) Time evolution of the observables $r_0(t),
r_1(t), r_2(t)$, and $r_3(t)$ for a single realization of the initial
state $\{\theta_i(0),\omega_i(0); i=1,2,\ldots,N\}$, where the $\theta_i$'s are chosen uniformly in
$[-\pi,\pi]$, while the $\omega_i$'s are chosen from a Gaussian
distribution with zero mean and unit variance, Eq. (\ref{Gaussian}).
Thus, initially, the system is in an unsynchronized state. Here, $\alpha=0.5$, and $K=15$. For this value of $\alpha$, Eqs. (\ref{Kcm}) and
(\ref{Gaussian}) give $K_c^{(0)} \approx 1.59577, K_c^{(1)} \approx
4.26696, K_c^{(2)} \approx 6.53664, K_c^{(3)} \approx 7.71516, \ldots$,
so that the Fourier modes $0,1,2,3$ are all linearly unstable.
Consequently, $r_0(t),
r_1(t), r_2(t)$, and $r_3(t)$ all grow in time from their initial values.  
The data in the plot are obtained from numerical simulations with
$N=2^{14}$.}
\l{fig1}
\end{figure}

\begin{figure*}[here!]
\centering
\centering
\begin{tabular}{lr}
\parbox[l]{9cm}{
\includegraphics[width=90mm]{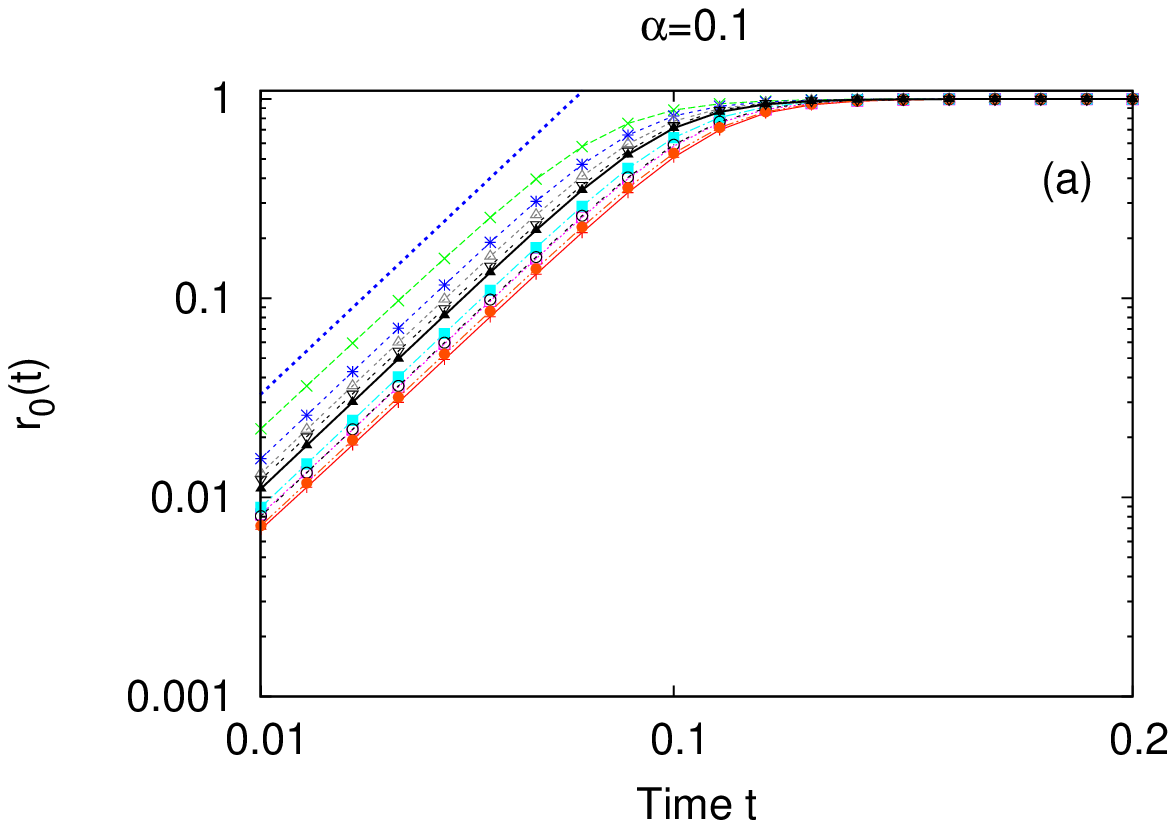}
}&
\parbox[r]{9cm}{
\includegraphics[width=90mm]{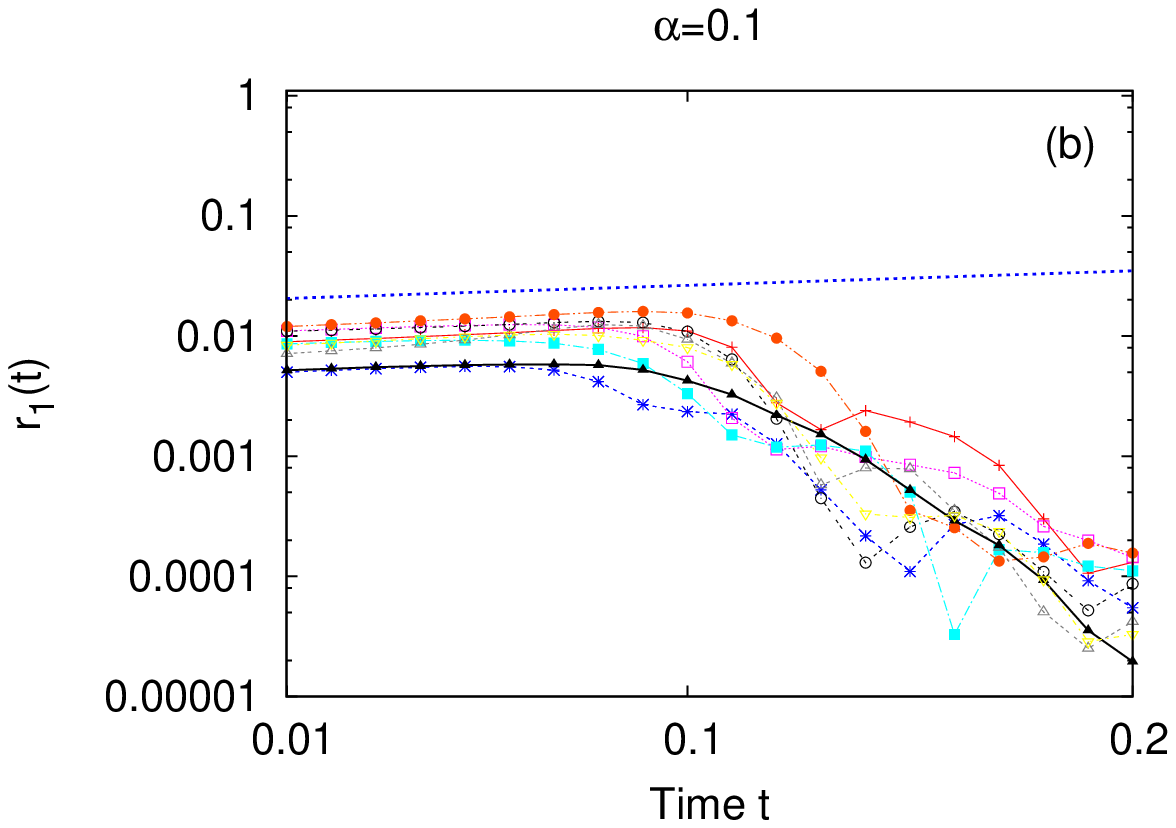}
}
\end{tabular}
\begin{tabular}{lr}
\parbox[l]{9cm}{
\includegraphics[width=90mm]{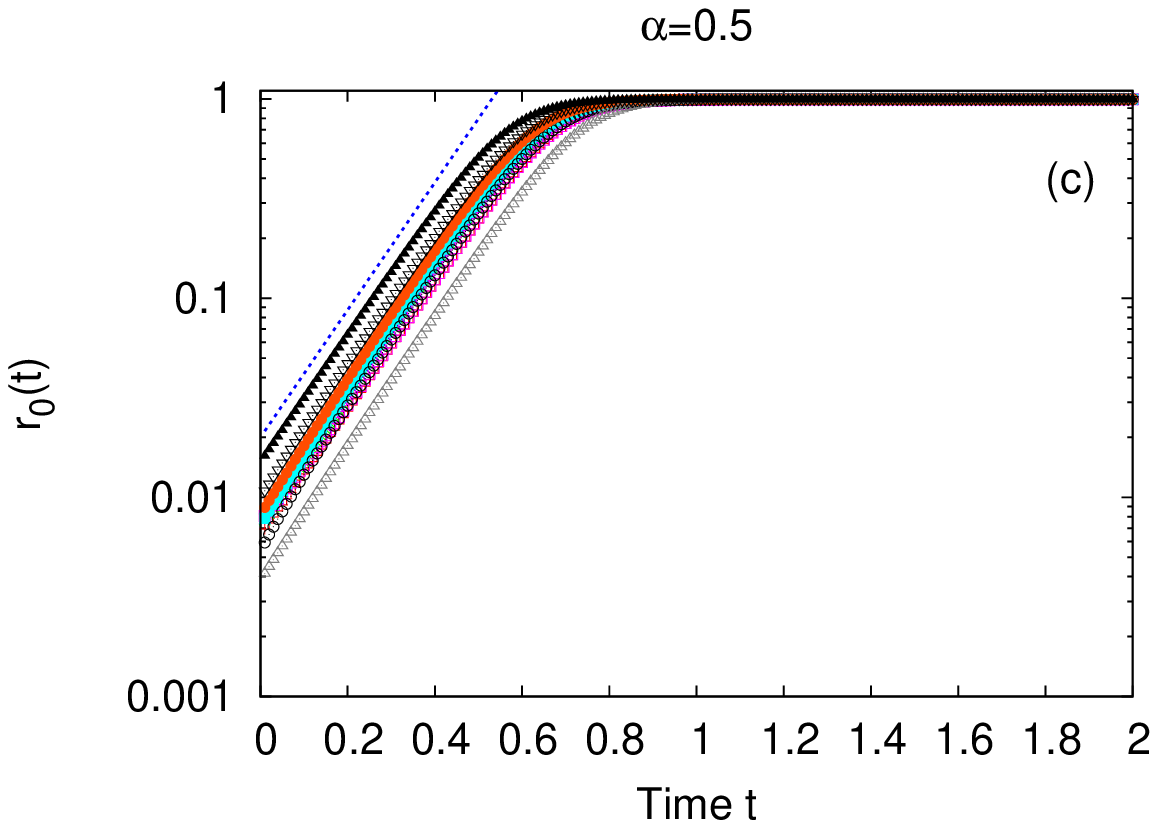}
}&
\parbox[r]{9cm}{
\includegraphics[width=90mm]{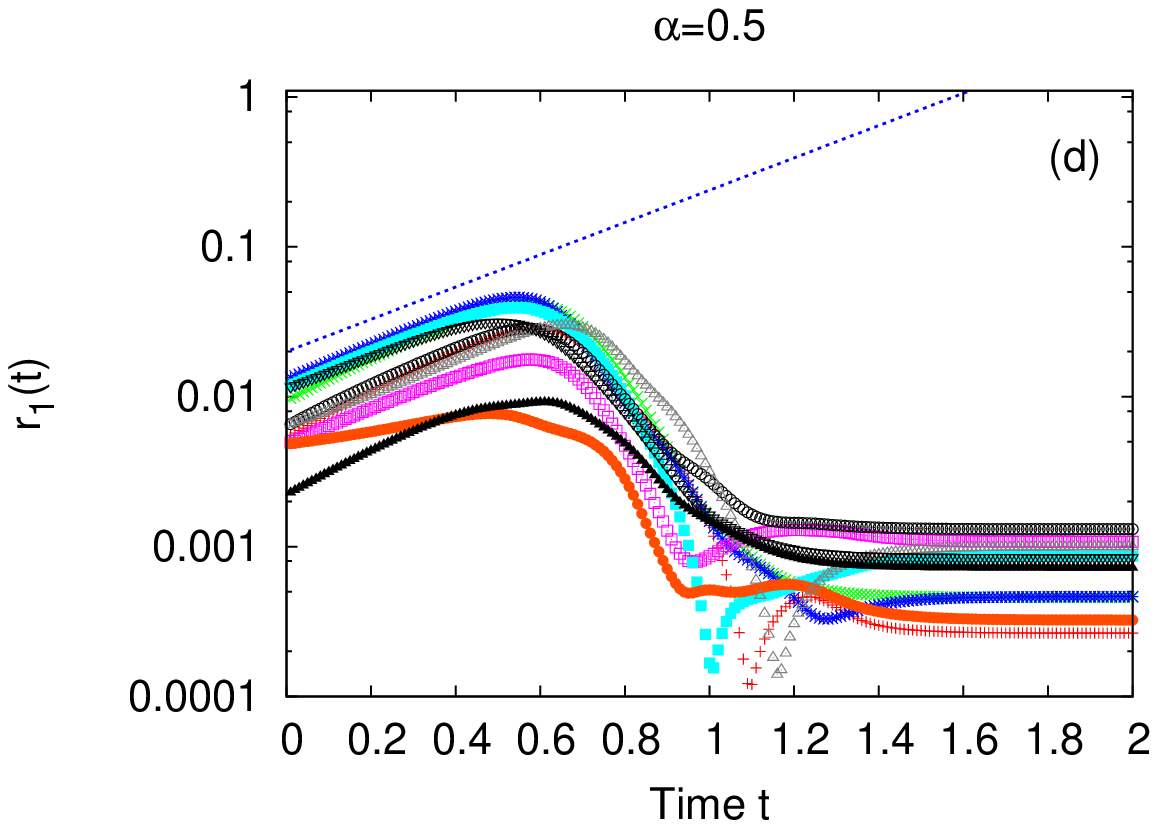}
}
\end{tabular}
\begin{tabular}{lr}
\parbox[l]{9cm}{
\includegraphics[width=90mm]{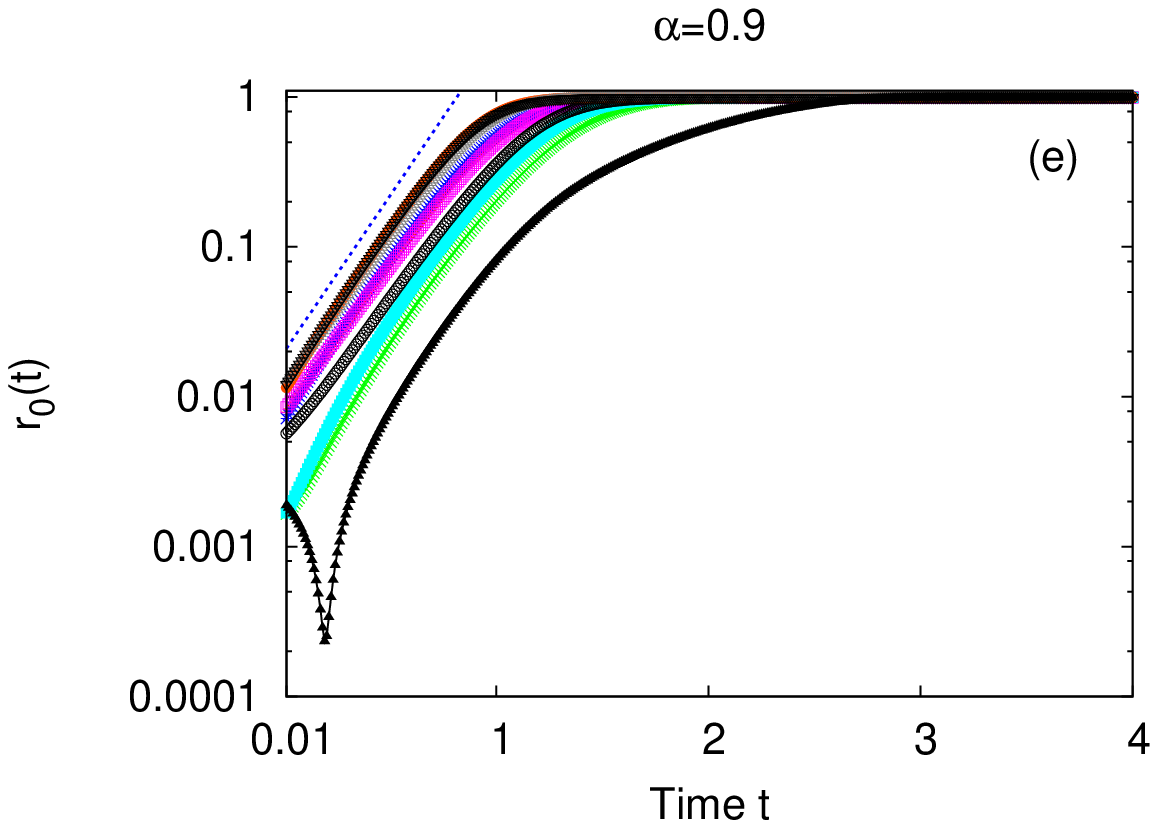}
}&
\parbox[r]{9cm}{
\includegraphics[width=90mm]{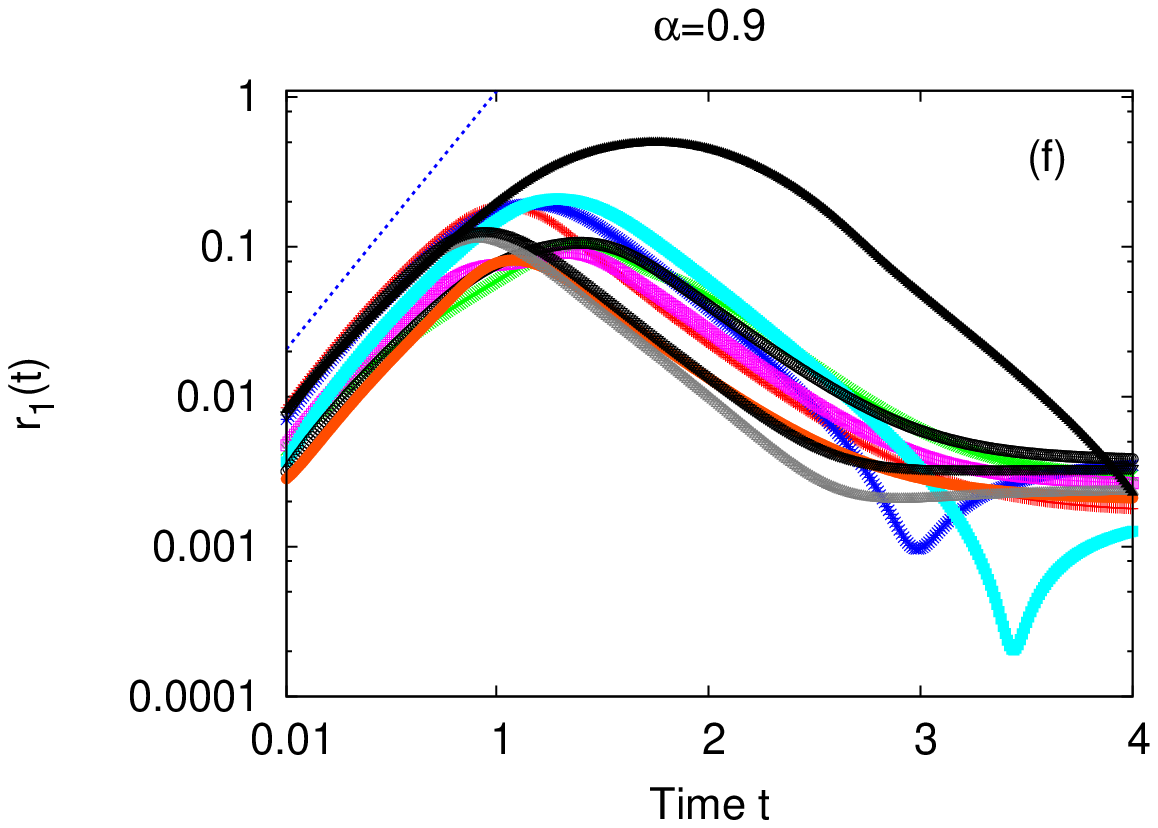}
}
\end{tabular}
\caption{(Color online) Time evolution of $r_0(t)$ and
$r_1(t)$ for $10$ different realizations of the initial
state $\{\theta_i(0),\omega_i(0);
i=1,2,\ldots,N\}$, where the $\theta_i$'s are chosen uniformly in
$[-\pi,\pi]$, while the $\omega_i$'s are chosen from a Gaussian
distribution with zero mean and unit variance, Eq. (\ref{Gaussian}).
Thus, initially, the system is in an unsynchronized state. Each figure
corresponds to a value of $\alpha$ indicated therein. For $\alpha=0.1$, we took
$K=100$. Equations (\ref{Kcm}) and
(\ref{Gaussian}) give $K_c^{(1)} \approx
25.8978, K_c^{(2)} \approx 59.9836, K_c^{(3)} \approx 76.1048, \ldots$. For $\alpha=0.5$, we took
$K=15$; here, $K_c^{(1)} \approx 4.26696, K_c^{(2)} \approx 6.53664,
K_c^{(3)} \approx 7.71516, \ldots$. For $\alpha=0.9$, we took
$K=10$; here, $K_c^{(1)} \approx 1.88898, K_c^{(2)} \approx 2.04629,
K_c^{(3)} \approx 2.12267, \ldots$. We have
$K_c^{(0)} \approx 1.59577$, independent of $\alpha$. 
Thus, for these three values of $\alpha$ with the corresponding values
of $K$, the Fourier modes $0$ and $1$ are linearly unstable. As a
result, in all cases, $r_0(t)$ and $r_1(t)$ grow in
time from their initial values as $r_0(t) \sim e^{\lambda_0 t}, r_1(t)
\sim e^{\lambda_1 t}$. 
The data in the plots are obtained from numerical simulations with
$N=2^{14}$. The dotted
blue line in each plot shows the exponential growth
with the rates $\lambda_0$ and $\lambda_1$ given by Eq.
(\ref{theoretical-growth}).}
\label{fig2}
\end{figure*}

\begin{figure}[h!]
\includegraphics[width=80mm]{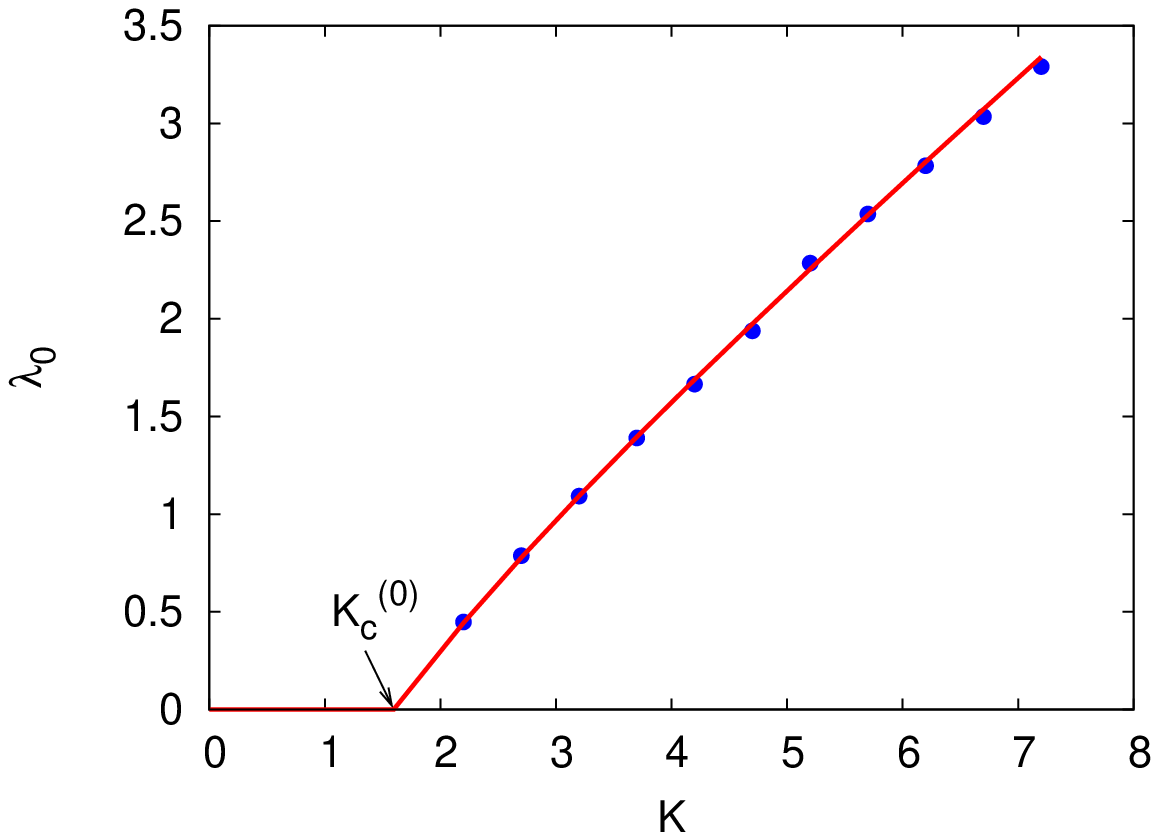}
\caption{(Color online) The points denote the values of the growth rate
$\lambda_0$ obtained by fitting the growth of $r_0(t)$ in time as
$r_0(t) \sim e^{\lambda_0 t}$ for short times, while starting from an
unsynchronized state with Gaussian $g(\o)$ (Eq. (\ref{Gaussian})). The
data for $r_0(t)$ were obtained by performing numerical simulations with $\alpha=0.5,
N=2^{14}$, and
by averaging over $100$ initial realizations. The numerical errors are of the size of the
points. The solid line shows the
theoretical curve for $\lambda_0$ given by Eq.
(\ref{theoretical-growth}). From our theoretical analysis, the growth
rate is zero for values of $K$ below $K_c^{(0)}$, and is non-zero above. In the figure, the
theoretical $K_c^{(0)}$ is indicated.}
\l{fig3}
\end{figure}

In order to have some representative numbers for the critical
thresholds $K_c^{(m)}$'s, and for comparison
with numerical simulations reported in the next section, let us choose a
Gaussian $g(\o)$ with zero mean and unit variance:
\be
g(\o)=\fr{1}{\sqrt{2\pi}}e^{-\o^2/2}.
\l{Gaussian}
\ee
Then, using Eq. (\ref{final-dispersion1}), we see that for
$K > K_c^{(m)}$, the growth rate $\lambda_m > 0$ of the $m$th mode satisfies
\be
\fr{K}{K_c^{(m)}}e^{\lambda_m^2/2}{\rm
Erfc}\Big[\fr{\lambda_m}{\sqrt{2}}\Big]=1,
\l{theoretical-growth}
\ee
where ${\rm Erfc}[x]$ is the complementary error function: ${\rm
Erfc}[x]=\fr{2}{\sqrt{\pi}}\int_x^\infty e^{-t^2}dt$.
In obtaining Eq. (\ref{theoretical-growth}), we have used the result $\int_{-\infty}^\infty dp \fr{e^{-\beta
p^2/2}}{p^2+a^2}=\fr{\pi}{a}e^{\beta
a^2/2}{\rm Erfc}\Big[\sqrt{\fr{\beta}{2}}a\Big]$. 

\section{Comparison with numerical simulations}
\l{simulations}

\begin{table}[h!]
\begin{tabular}{|c|c|c|c|c|c|}
\hline
 $\alpha$ & $K_c^{(1)}$ & $K_c^{(2)}$ & $K_c^{(3)}$ & $K_c^{(4)}$ &
 $K_c^{(5)}$   \\ \hline \hline 
0.05 & 52.96519  & 130.12769 & 165.38633 & 240.09713 & 274.75919   \\ \hline 
0.1 & 25.89784 & 59.98356 & 76.10483 & 107.39521 &122.97195  \\ \hline 
0.5 & 4.26696 & 6.53664 & 7.71516 & 9.10681 & 10.02918  \\ \hline 
0.9 & 1.88898 & 2.04629 & 2.12267 & 2.18907 &2.23565   \\ \hline 
0.95 & 1.73443 & 1.80472 & 1.83837 & 1.86673 & 1.88661  \\ \hline 
\end{tabular}
\caption{$K_c^{(m)}; m=1,2,\ldots,5$ as a function of $\alpha$ for the
frequency distribution (\ref{Gaussian}). In all
cases, $K_c^{(0)} \approx 1.59577$.}
\l{table1}
\end{table}

Here, we test the theoretical predictions of the preceding
Section by comparing them with results from numerical simulations. A standard procedure for
simulations is to integrate the equations of motion (\ref{EOM}). 
However, note that the equations involve computing at every integration step a sum over $N$ terms
for each of the $N$ oscillators, and therefore require a
total computation time that would scale with $N$ as $N^2$. In the Appendix
\ref{app1}, we show that the equations of motion can be
transformed into a convenient form which when integrated requires at
every integration step a computation time that scales with $N$ as $N \ln
N$. We chose this alternative form of the equations of
motion, and integrated them using a fourth-order Runge-Kutta algorithm with
time step equal to $0.01$. The initial state of the system was chosen to have each
oscillator phase $\th$ uniformly distributed in $[-\pi,\pi]$ and frequency $\o$ sampled from the Gaussian distribution
(\ref{Gaussian}), independently of all the other oscillators. Thus, initially, the system is in an unsynchronized state. We report here simulation results for $N=2^{14}$.

We specifically compare the theoretical and simulation results for the
growth rates $\lambda_m$'s. From Eq. (\ref{theoretical-growth}), these rates depend on the
critical thresholds $K_c^{(m)}$'s, and, thus, their comparison would
indirectly provide a test of the theoretical predictions for
$K_c^{(m)}$'s given by Eqs. (\ref{Kcm}) and (\ref{Gaussian}) as
\be
K_c^{(m)}=\fr{2\sqrt{2}}{\sqrt{\pi}\kappa(\a)\Lambda_m(\a)}.
\ee

Using the above equation, we show in Table \ref{table1} the dependence of $K_c^{(m)}$'s for
$m=1,2,3,4,5$, as a function of $\alpha$; note that $K_c^{(0)}$ is independent of
$\alpha$.
From the table, we see that the
thresholds for the non-zero Fourier modes get larger and further apart
when $\alpha$ is close to zero. On the contrary, these thresholds approach the
threshold for the zero mode when $\alpha$ has a value close to unity. From this
observation, one may guess that in the limit of $\alpha$ approaching zero, all thresholds
excepting the one for the zero mode become infinitely large, while in
the limit of $\alpha$ approaching unity, they all acquire the common value of the
zero-mode threshold.

In simulations, we monitor the time evolution of the
observable
\be
r_m(t)=\fr{1}{N}\Big|\sum_{j=1}^N e^{i(\th_j+2\pi jm/N)}\Big|;
m=0,1,2,\ldots.
\ee
Note that $r_0(t)$ corresponds to the case in which there is no spatial dependence of the oscillator phases, and, thus, characterizes the
mean-field mode.  

From our analysis in the preceding Section, we see that for $K >
K_c^{(m)}$, when the $m$th spatial mode of fluctuations is
linearly unstable, $r_m(t)$ should grow exponentially in time: $r_m(t) \sim
e^{\lambda_m t}$. In our simulations, we found that for large enough $K$,
the observable $r_0(t)$ grows in time and saturates to a value of
$O(1)$, while all the other $r_m(t)$'s, after showing an initial growth
in time, decay to a value close to zero. An
illustration of such a behavior is shown in Fig. \ref{fig1}. This
observation implies that in the long-time limit, the dynamics is dominated by the
mean-field mode, which drives the system to a synchronized state
signaled by $r_0(t)$ settling to a value of order $1$.

In Fig. \ref{fig2}, we show the temporal evolution of $r_0(t)$ and
$r_1(t)$ for $10$ different realizations of the initial state
for three different values of $\a$, namely, a value close to zero, a value
close to unity, and a value in between. In each case, the value of $K$ is such
that the Fourier modes $0$ and $1$ are linearly unstable; thus, $K$ is
at least larger than $K_c^{(1)}$.
Consequently, $r_0(t)$ and $r_1(t)$ are expected to grow in
time from their initial values as $r_0(t) \sim e^{\lambda_0 t}, r_1(t)
\sim e^{\lambda_1 t}$. 

One may see from the plots for $r_0(t)$ in Fig. \ref{fig2} that almost all realizations
exhibit growth rates close to the theoretical value. It is then natural
to average $r_0(t)$ over large enough number of initial realizations in order to
reduce statistical fluctuations, and extract the growth rate
$\lambda_0$. We perform this exercise for $\a=0.5$, and
display the result for $\lambda_0$ as a function of $K$ in Fig.
\ref{fig3}, which shows a very good agreement with the theoretical prediction.  

From the plots of $r_1(t)$ in Fig. \ref{fig2}, one may note that the
growth rates for most initial realizations are close to the theoretical
value, although there are some realizations for which significant
deviations are seen. On decreasing the value of $K$ towards $K_c^{(1)}$, the number of this
latter class of realizations increases, most significantly when $K$ is
close to $K_c^{(1)}$. We believe that this
could be due to pronounced finite-$N$ effects close to $K_c^{(1)}$. Also,
note from Eq. (\ref{theoretical-growth}) that $\lambda_1$ decreases with
decreasing $K$. Since $r_1(t)$ in the limit of long times decays to a
close-to-zero value, it
leaves with a very narrow time window for the growth of $r_1(t)$.

From the plots in Fig. \ref{fig2}, one may observe that the growth rate
of $r_1(t)$ goes to zero when $\a$ approaches zero, while it gets close
to the rate for $r_0(t)$ when $\a$ approaches unity. This is in accordance
with our discussion above on the behavior of thresholds as $\a$
approaches zero and unity.

\section{Concluding remarks}
In this paper, we considered a system of coupled phase-only oscillators, in
which each site of a one-dimensional periodic lattice of $N$ sites
contains one oscillator of distributed natural frequency. The oscillators are coupled
through a power-law interaction $\sim (K /\widetilde{N})r^{-\a}$ in their separation $r$
along the lattice length, where the exponent $\a$ lies in the range
$0 \le \a <1$, and $\widetilde{N}$ is a size-dependent normalization. The oscillator frequencies are taken to be distributed according
to a density $g(\o)$ that is symmetric about its mean (taken to be $0$) and non-increasing on $[0,\infty)$.

We studied the model in the continuum limit $N \to \infty$. In this
limit, one may define a local density of oscillators which evolves in time
according to the continuity equation expressing the conservation of
number of oscillators of each frequency under the dynamics. The unsynchronized state, uniform both in phase and in spatial coordinate, represents a stationary solution of the continuity equation. By
analyzing the linear stability of such a state, we showed that when it is
unstable, different spatial Fourier modes of fluctuations have different
stability thresholds beyond which they grow exponentially in
time with different rates. However, numerical simulations show that at long times, all the non-zero
Fourier modes decay in time, while the zero mode (the ``mean-field"
mode) grows in time to dominate the instability process and drive the system to a synchronized state.
Such a long-time dominance of the
mean-field mode is known for systems with Hamiltonian systems
\cite{Bachelard:2011}. The present
work illustrates the dominance for dissipative dynamics within the ambit
of a simple model, while leaving the proof of its general validity as an interesting
open issue.

\section{Acknowledgments}
We acknowledge useful discussions with Romain Bachelard and Thierry Dauxois, and financial support from INFN and ANR-10-CEXC-010-01, Chaire d'Excellence
project. We are especially grateful to Alessandro Campa for many
fruitful discussions, valuable inputs, and a careful reading of the
manuscript. We thank Arkady Pikovsky for pointing out Ref.
\cite{Pikovsky:2011} to us. S. G. thanks Anandamohan Ghosh for directing
him to the Kuramoto
model.
\appendix

\section{A fast numerical algorithm to compute the interaction term in Eq. (\ref{EOM})}
\l{app1}
Here we discuss a numerical algorithm that transforms the equations of
motion (\ref{EOM}) into a convenient form in which the interaction term
is efficiently computed by a Fast Fourier Transform
(FFT) scheme in
times of order $N \ln N$ for a system of $N$ oscillators. Use of FFT
requires that we choose a power of $2$ for $N$. 

We start with Eq. (\ref{EOM}) which we rewrite as
\be
\dot{\th_{i}}=\omega_{i}+\frac{K}{\widetilde{N}}\sum_{j=1}^{N}\fr{\sin(\th_j-\th_i)}{(d_{ij})^{\a}},
\l{EOM1}
\ee
where $d_{ij}$ is the shortest distance between sites $i$ and $j$ on a
periodic lattice of $N$ sites. We
set $d_{ii}=0$, while $d_{ij}$ for $i \ne j$ is given by
\be
d_{ij}=\left\{ 
\begin{array}{ll}
                |j-i| & \mbox{if $1 \le |j-i| \le N/2$}, \\
               N-|j-i| & \mbox{otherwise}.
               \end{array}
        \right. \\
\label{dij}
\ee

Equation (\ref{EOM1}) may be rewritten as
\be
\dot{\th_{i}}=\omega_{i}+\frac{K}{\widetilde{N}}\Big(\cos \th_i
\sum_{j=1}^{N}V_{ij}\sin\th_j- \sin \th_i \sum_{j=1}^{N}V_{ij} \cos
\th_j\Big).
\l{EOM2}
\ee
Here, the first summation term may be interpreted as the $i$th element
of the $N \times 1$ column vector formed by the product of an $N \times N$ matrix $V=[V_{ij}]_{i,j=1,2,\ldots,N}$ with an $N \times 1$ column
vector $(\sin \th_1 \sin \th_2 \ldots, \sin \th_N)^T$, where $V_{ij}=1/(d_{ij})^\a$.
Similarly, the second summation term may be interpreted as the $i$th element
of the $N \times 1$ column vector formed by the product of $V$ with an $N \times 1$ column vector $(\cos \th_1
\cos \th_2 \ldots, \cos \th_N)^T$.

Writing out the matrix $V$, we see that it has the form
\bea
V = \begin{bmatrix}
v_1     & v_{N} & \dots  & v_{3} & v_{2}  \\
v_{2} & v_1    & v_{N} &         & v_{3}  \\
\vdots  & v_{2}& v_1    & \ddots  & \vdots   \\
v_{N-1}  &        & \ddots & \ddots  & v_{N}   \\
v_{N}  & v_{N-1} & \dots  & v_{2} & v_1 \\
\end{bmatrix}, 
\eea
where $v_1=0$ and 
\be
v_r=\left\{ 
\begin{array}{ll}
                1/(r-1)^\a & \mbox{if $2 \le r \le N/2+1$}, \\
               1/(N-r+1)^\a & \mbox{otherwise}.
               \end{array}
        \right. \\
\label{vr}
\ee
Thus, $V$ is a circulant matrix fully specified by the elements in the
first column. The remaining columns of $V$ are each cyclic permutations of the elements
in the first column, with offset equal to the column index.

Now, note that $V$ can be written as 
\be
V=v_1 I+ v_2 P + v_3 P^2+\ldots+v_{N}P^{N-1},
\l{VP}
\ee
where $P$ is the $N \times N$ cyclic permutation matrix,
\be
\begin{bmatrix}
0&0&\ldots&0&1\\
 1&0&\ldots&0&0\\
 0&\ddots&\ddots&\vdots&\vdots\\
 \vdots&\ddots&\ddots&0&0\\
 0&\ldots&0&1&0
\end{bmatrix}
.
\ee
Since $P^N=I$, the $N \times N$ identity matrix, the eigenvalues of $P$
are given by $w_j=e^{i2\pi (j-1)/N}; j=1,2,\ldots,N$, where $w_j$ is the
$N$-th root of unity.
Equation (\ref{VP}) then implies that the eigenvalues of $V$ are given by $\Lambda_j=\sum_{k=1}^{N}v_k
w_j^{k-1}$ for $j=1,2,\ldots,N$.

One may easily check that the
eigenvectors of $V$ are the columns of the $N \times N$ unitary discrete
Fourier transform matrix
$F=\fr{1}{\sqrt{N}}[f_{jk}]_{j,k=1,2,\ldots,N}$, where
\be
f_{jk}=e^{-i2\pi (j-1)(k-1)/N} {\rm ~for~} 1 \le j,k \le N.
\ee
Then, one has $F^{-1}VF=[\Lambda_j \delta_{ij}]_{i,j=1,2,\ldots,N}$. 

In
terms of the matrices $F$ and $F^{-1}$, one may easily rewrite Eq.
(\ref{EOM2}) as
\bea
\dot{\th_{i}}&=&\omega_{i}+\fr{K}{\widetilde{N}}\Big[\cos \th_i \sum_{j=1}^N
(F^{-1})_{ij}\Lambda_j (F \sin \th)_j\nonumber \\
&&-\sin \th_i \sum_{j=1}^N (F^{-1})_{ij}\Lambda_j (F \cos \th)_j\Big],
\l{EOM3}
\eea
where $(F \sin \th)_j$ (respectively, $(F \cos \th)_j$) is the $j$th element of the $N \times 1$ column vector formed by
multiplying the matrix $F$ with the column vector $(\sin \th_1
\sin \th_2 \ldots \sin \th_N)^T$ (respectively, $(\cos \th_1 \cos \th_2
\ldots \cos \th_N)^T$). Note that $(F \sin \th)_j$ and $(F \cos \th)_j$
are just discrete Fourier transforms, and may be computed very
efficiently by standard FFT codes (see, e.g., Ref. \cite{Antia}). The simulations reported in Section
\ref{simulations} were performed by integrating Eq. (\ref{EOM3}).

\end{document}